# Comparing Terminal Performance of .357 SIG and 9mm Bullets in Ballistic Gelatin Using Retarding Force Analysis from High Speed Video

Elizabeth Keys, Amy Courtney, and Michael Courtney

Michael_Courtney@alum.mit.edu

**Abstract**

High-speed video has emerged as an valuable tool for quantifying bullet performance in ballistic gelatin. This paper presents the results of testing four .357 SIG bullets using high-speed video of bullet impacts in ballistic gelatin to determine retarding force curves, permanent cavities, temporary cavities, and energy deposit vs. penetration depth. Since the methods are identical, results are meaningfully compared with four 9mm NATO bullets studied in an earlier project. Though .357 SIG bullets perform slightly better due to higher impact energy, the principal finding is that there is a much bigger difference in performance between the best and worst performing bullets in each cartridge than there is between bullets of similar design in the two cartridges. In each cartridge, higher performing expanding bullets (jacketed hollow points) outperform non-expanding bullets (full metal jacket) by a wide margin, showing a much higher probability of rapid incapacitation according to an Army Research Laboratory model by Neades and Prather (1991).

**Introduction**

Quantifying bullet performance is a necessary step in the ammunition selection process. Three physical mechanisms have been shown to contribute to the wounding and incapacitation potential of penetrating projectiles: the permanent cavity, the temporary cavity, and a ballistic pressure wave radiating outward from the projectile (Courtney and Courtney, 2012). All three physical mechanisms originate in the retarding force between bullet and tissue (Peters, 1990). The permanent cavity is often described as being caused by direct contact between bullet and tissue, but it is actually caused by an intense stress field in the immediate vicinity of the passing projectile. These stress waves decay rapidly with distance to levels below tissue damage thresholds a short distance from the path of the bullet through tissue. The temporary cavity is caused by the retarding force accelerating tissue forward and outward from the bullet path until elasticity causes the tissue to spring back into place. Stretch beyond tissue's elastic limit can enlarge the permanent cavity. Stretching can also cause nerve damage, and impact of the temporary stretch cavity with the spine can cause injury and hasten incapacitation. A ballistic pressure wave travels outward from the bullet path through tissue and can also contribute to wounding and incapacitation (Courtney and Courtney, 2007a; Suneson et al. 1990a, 1990b; Krasja, 2009; Selman et al., 2011).

Calibrated 10% ballistic gelatin has become widely accepted as a homogeneous tissue simulant that adequately reproduces average retarding forces and penetration depths of projectiles in tissue. It is common to use ballistic gelatin to quantify penetration depth, permanent cavity, and temporary cavity volumes, and it has recently been demonstrated that high speed video can be used to quantify the retarding force between bullet and tissue (Gaylord, 2013). Bo Janzon and colleagues seem to have originated the work of quantifying the retarding force and emphasizing its importance in wound ballistics (Janzon, 1983; Bellamy and Zajtchuk, 1990).

Since deformation of the target is occurring, it is understood that work is being done on the target by the projectile. The energy lost by the projectile can be related to the force of the projectile by the Work-Energy Theorem. Newton's Third Law of Motion states the force applied by the bullet on the target is equal and opposite the force of the target on the bullet. Because of this equivalence, the retarding force of the target, which is equal to the force of the bullet on the target, can be obtained.

# Comparing Terminal Performance of .357 SIG and 9mm Bullets in Ballistic Gelatin Using Retarding Force Analysis from High Speed Video

Examination of ballistic gelatin by direct observation cannot occur in real-time as the projectile passes too quickly. In years past, micro-flash or x-ray flash photography was used to assess fragmentation characteristics and analyze the flight path of projectiles (Barber, 1956). The use of an accurate and readily available technology in the form of a high-speed camera is employed here. The path of the projectile is recorded beginning just prior to impact until the projectile comes to rest or exits the gelatin. The position of the projectile can be tracked. With an appropriate scale in the video, pixels can be converted to length, and then into a change of distance traveled by the bullet between the frames of the video. This computation yields velocity of the projectile vs. time from which acceleration can be calculated. The change in kinetic energy can also be computed. Temporary and permanent cavities are also visible. In this report, four .357 SIG rounds will be analyzed using the high-speed video method showing retarding force versus penetration depth, and the results will be compared with earlier work (Gaylord et al., 2013) quantifying bullet performance in four 9mm NATO loads.

**Method**
The experimental and analysis methods are the same as previously described in Gaylord et al. (2013), except that four factory .357 SIG loads were used in place of the four 9mm NATO loads in the original study. These projectiles included a 125 grain Full Metal Jacket (125 FMJ), a 115 grain Speer Gold Dot Hollow Point (115 GDHP), a 125 grain Federal Premium Personal Defense Hollow Point (125 FedHP), and a 125 grain Winchester Ranger SXT (125 SXT). Rounds were fired from an appropriately chambered SIG P229 pistol.

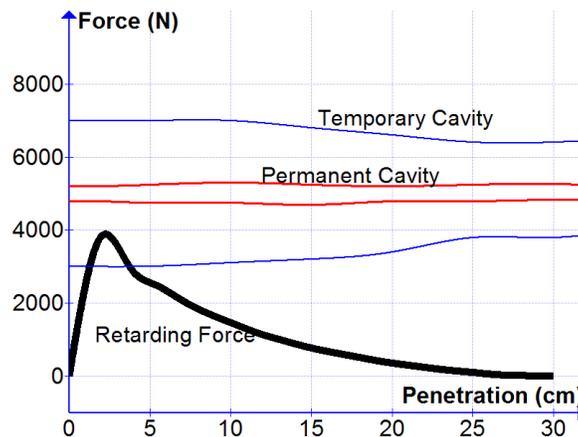

Figure 1: Retarding force vs. penetration depth for the .357 SIG 125 grain FMJ. The permanent and temporary cavities are also shown. The vertical and horizontal length scales are the same.

**Results**
Figure 1 shows the force curve, temporary cavity, and permanent cavity of the 125 grain FMJ in .357 SIG. Compared with the 124 grain FMJ in 9mm NATO (Figure 4 of Gaylord et al., 2013), the .357 SIG load has a peak retarding force close to 4000 N, rather than 2100 N. Both temporary cavities have a maximum diameter close to 10cm, but the .357 SIG FMJ bullet reaches the peak diameter quickly and then the diameter decreases; whereas, both the retarding force and the temporary cavity oscillate for the 9 mm NATO FMJ. This is because the .357 SIG bullet remains point forward while penetrating the gelatin; whereas, the 9mm NATO bullet tumbles. The peaks in the 9mm NATO temporary cavity diameter and retarding force curves occur when the bullet presents a larger cross sectional area by traveling sideways.

In addition to higher impact velocity (1309 ft/s for the .357 SIG vs. 1156 ft/s for the 9mm NATO FMJ), the bullet shape shown in Figure 2 helps explain the differences in bullet performance. Because the .357 SIG cartridge uses a bottleneck design, its FMJ bullet can have a flat nose profile without causing feed-ing issues in semiautomatic pistols. This flatter nose profile creates a wider temporary cavity and a higher peak retarding force than the round nose profile of the 9mm NATO, and tumbling is not necessary to reach the peak temporary cavity diameter or retarding force.

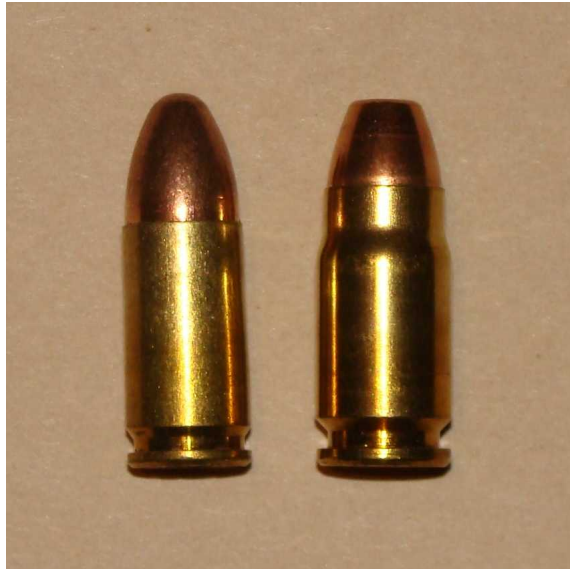

Figure 2: Loaded cartridges for the .357 SIG 125 FMJ (right) and the 9mm NATO 124 FMJ (left).

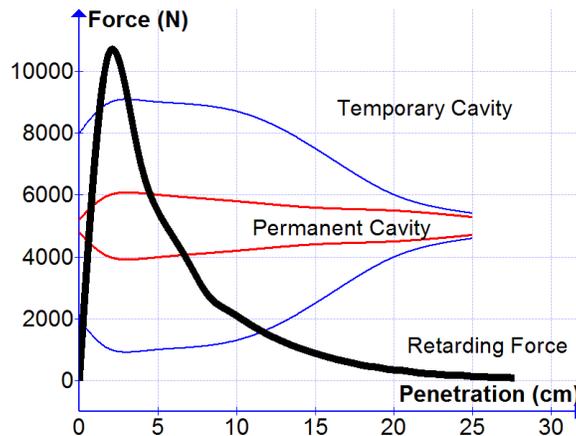

Figure 3: Retarding force vs. penetration depth for the .357 SIG 125 SXT load. Permanent and temporary cavities are also shown.

Figure 3 shows the force curve, temporary cavity, and permanent cavity of the 125 grain SXT in .357 SIG. This jacketed hollow point expanding bullet has a much larger peak retarding force, temporary cavity diameter, and permanent cavity diameter than non-expanding FMJ bullets. At over 10,000 N, the peak retarding force is also larger than what has been measured for other bullets in .357 SIG and 9mm NATO. The bullet jacket is a thick copper alloy which is intentionally scored and engineered to open quickly revealing six sharp edges. (The design was once marketed as the Winchester Black Talon.) However, after reaching a maximum expanded diameter, the sharp petals continue to fold back toward the bullet base reducing the effective cross sectional area. The final spape is shown in Figure 4. While the bullet is at or near maximum diameter, there is likely a cutting effect contributing to wounding, but after the petals fold backward, the permanent cavity is likely formed mostly by direct crushing and the "prompt damage" of the high intensity stress field (Peters, 1990).

# Comparing Terminal Performance of .357 SIG and 9mm Bullets in Ballistic Gelatin Using Retarding Force Analysis from High Speed Video

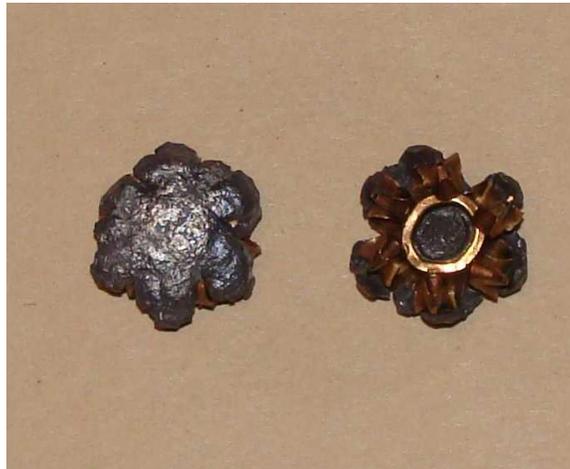

*Figure 4: Front and back of expanded .357 SIG 125 SXT bullet after recovery from ballistic gelatin.*

The temporary cavity shown in Figure 3 for the 125 SXT reaches full diameter at a relatively shallow penetration depth of 2 cm. The expanded diameter of 0.775" explains why. Such a large amount of early expansion likely gives the tissue simulant a much greater transverse acceleration than bullets that expand to a smaller diameter, such as the 9mm NATO 127 SXT which only expands to 0.550".

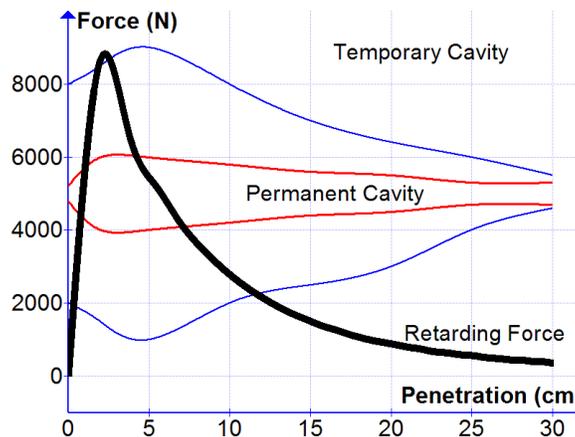

*Figure 5: Retarding force vs. penetration depth for the .357 SIG 115 GDHP load. Permanent and temporary cavities are also shown.*

The force curve, temporary cavity, and permanent cavity for the 115 GDHP load in .357 SIG are shown in Figure 5. This light weight and fast bullet impacts at 1489 ft/s resulting in very rapid expansion, a peak retarding force over 8500 N at a penetration depth of 2 cm, and a large diameter temporary cavity. However, neither the large retarding force nor the large diameter temporary cavity are sustained at deeper penetration depths. Unlike typical jacketed lead handgun bullets, this bullet design has the jacket strongly bonded to the core by an electroplating process that builds up a thick layer of copper-zinc alloy around the soft lead core. This process results in a stout bullet that resists fragmentation and retains mass even at high impact velocities.

# Comparing Terminal Performance of .357 SIG and 9mm Bullets in Ballistic Gelatin Using Retarding Force Analysis from High Speed Video

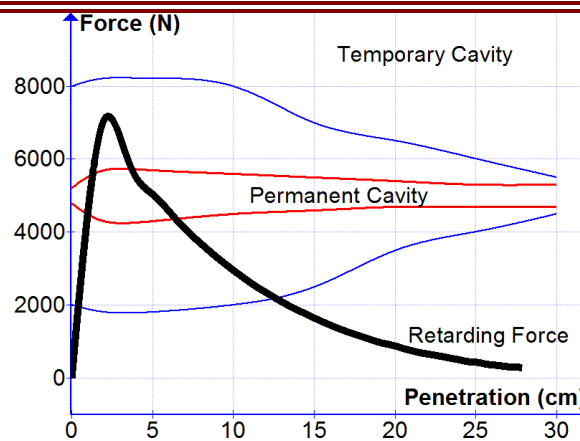

*Figure 6: Retarding force vs. penetration depth for the .357 SIG 125 FedHP load. Permanent and temporary cavities are also shown.*

Figure 6 shows the force curve, temporary cavity, and permanent cavity of the 125 grain Fed HP in .357 SIG. The peak retarding force is just over 7000 N. This bullet from the Federal Premium personal defense line is a conventional jacketed lead hollow point with a jacket that is not bonded to the lead core, but is scored in the frontal area to assist with uniform expansion and penetration.

This video analysis method can be applied to determine the energy deposit in any desired range of penetration depths. Bo Janzon used energy deposit to estimate the volume/diameter of wounded tissue as a function of penetration depth (Janzon, 1983). Neades and Prather (1991) published a method to estimate the probability of incapacitation given a hit, P(I/H), on an enemy soldier based on the energy deposit in the first 15 cm of penetration (E15). Implementation details for handgun bullets in gelatin are described in Gaylord et al. (2013).

| Load | E15 | P(I/H) | M193 Equivalent Range |
|---|---|---|---|
| **9mm NATO** | (ft lbs) | | (yards) |
| 124 FMJ | 172 | 0.330 | 620 |
| 147 WWB | 197 | 0.361 | 570 |
| 147 SXT | 240 | 0.410 | 490 |
| 127 SXT | 350 | 0.515 | 380 |
| **.357SIG** | | | |
| 125 FMJ | 222 | 0.390 | 530 |
| 125 SXT | 466 | 0.600 | 270 |
| 115 GDHP | 467 | 0.601 | 270 |
| 125 FedHP | 437 | 0.581 | 290 |

*Table 1: Conditional incapacitation probabilities (PI/H) for .357 SIG and 9mm NATO bullets using the BRL method (Bruchey and Sturdivan, 1968; Neades and Prather, 1991). The range of equivalent effectiveness for the M16 (using the M193 bullet) is also shown (M16, 1968).*

Table 1 shows the resulting conditional incapacitation probabilities for the four .357 SIG loads in the present study compared with the four 9mm NATO loads from Gaylord et al. (2013). In .357 SIG, the FMJ bullet has a higher energy transfer in the first 15 cm than the FMJ bullet in 9mm NATO. This leads to a higher estimate for the conditional incapacitation probability. Furthermore, all three of the hollow point bullets in .357 SIG have higher E15 and thus higher incapacitation probabilities than any of the bullets

# Comparing Terminal Performance of .357 SIG and 9mm Bullets in Ballistic Gelatin Using Retarding Force Analysis from High Speed Video

tested in 9mm NATO. From this result, one might conclude that the .357 SIG generally provides more wounding potential than the 9mm NATO.

## Discussion

The 125 grain FMJ in .357 SIG bullet has a P(I/H) of only 39%, approximately the same as the M193 bullet at 530 yards fired from an M16 rifle. The M16 family of rifles is rarely effective beyond 300 yards (Ehrhart, 2009). Considering that pistols are commonly employed in close quarters, such a lack of effectiveness is unacceptable. The expected poor performance is consistent with the poor performance of all FMJ and other non-expanding handgun bullets in the Marshall and Sanow (2001) epidemiological type analysis of shootings in humans as well as the long incapacitation times resulting from use of all FMJ and non-expanding handgun bullets in a laboratory study in live goats (Courtney and Courtney, 2007c) .

All three jacketed hollow point rounds in .357 SIG performed better than all four 9mm NATO bullets from the previous study (Gaylord, et al., 2013). The 125 grain SXT in .357 SIG has a P(I/H) of 60%, corresponding to an effectiveness comparable to the M193 bullet (M16) at 270 yards. Handguns may never offer the terminal effectiveness of rifles at close distances, but this load may approximate near maximum performance available in pistols with controllable recoil. This load was unavailable when earlier studies were conducted that directly assessed pistol bullet effectiveness in humans (Marshall and Sanow, 2001) and goats (Courtney and Courtney, 2007c), but the bullet weight, design, and performance in ballistic gelatin are all very similar to the best performing JHP pistol bullets in those studies.

The 115 grain GDHP in .357 SIG also has a conditional incapacitation probability of 60%, approximating the terminal effectiveness of the M193 bullet (M16) at 270 yards. However, its peak retarding force is smaller than that of the 125 grain SXT in .357 SIG, as are the expanded bullet diameter and temporary cavity volume. Though the energies transferred in the first 15 cm of penetration are nearly identical, the differences in wound profile suggest that the 125 grain SXT has more potential for wounding of vital organs.

The 125 grain FedHP in .357 SIG has a P(I/H) of 58%, corresponding to an effectiveness comparable to the M193 bullet (M16) at 290 yards. One of the authors (MC) has also shot several whitetail deer with this load (handgun hunting) and observed impressive terminal effects both in the internal wounding as well as short distance run before collapse (25-50 yards). Another author (AC) carried this round in a handgun for many years and can report that it is both effective on various varmints around the homestead as well as accurate and controllable for a woman of average size and strength in both field use and extensive training.

In summary, this paper presents results using analysis of high speed video to quantify retarding forces and wound cavities of four .357 SIG loads in calibrated ballistic gelatin. In all cases, .357 SIG loads produce more impressive wound channels in tissue simulant than comparable 9mm NATO loads. Further, expanding bullets produce greater wounding effectiveness than non-expanding bullets from the same cartridge.

## Acknowledgements

This research was supported in part by BTG Research (www.btgresearch.org) and by the United States Air Force Academy. The views expressed in this paper are those of the authors and do not necessarily represent those of the U.S. Air Force Academy, the U.S. Air Force, the Department of Defense, or the U.S. Government. E. Keys wishes to acknowledge the assistance of Dr. Gary Shoemaker, Emeritus Faculty, Department of Physics and Astronomy California State University, Sacramento for invaluable help in error analysis, and to Christopher R. Keys, B. Sc. Physics (2015) for helpful discussions.

# Comparing Terminal Performance of .357 SIG and 9mm Bullets in Ballistic Gelatin Using Retarding Force Analysis from High Speed Video

# Comparing Terminal Performance of .357 SIG and 9mm Bullets in Ballistic Gelatin Using Retarding Force Analysis from High Speed Video

Elizabeth Keys
Department of Physics and Astronomy
California State University
6000 J$^{st}$
Sacramento, California, 95819
Elizabeth.I.Keys@gmail.com

Amy Courtney
Exponent, Inc.
3440 Market Street
Philadelphia, PA 19104, USA
Amy_Courtney@post.harvard.edu

Michael Courtney
BTG Research
Baton Rouge, Louisiana
Michael_Courtney@alum.mit.edu